\documentclass[9pt]{olplainarticle}
\usepackage{textgreek,graphicx}
\usepackage{rotating}
\usepackage{float}
\usepackage{xr}
\usepackage{hyperref}
\usepackage[normalem]{ulem}
\usepackage{stmaryrd}
\expandafter\def\csname opt@stmaryrd.sty\endcsname
{only,shortleftarrow,shortrightarrow}
\usepackage{extpfeil}
\usepackage{ctable} 

\definecolor{mycolor}{RGB}{0, 0, 0}
\usepackage{moreverb} 

\immediate\write18{texcount -inc -incbib 
-sum main_text.tex > /tmp/wordcount.tex}

\title{Therapeutic Interfering Particles Exploiting Viral Replication and Assembly Mechanisms Show Promising Performance: A Modelling Study}

\author[1,2,+]{Farzad Fatehi}
\author[1,2,3,+]{Richard J. Bingham}
\author[1,2,4,+]{Pierre-Philippe Dechant}
\author[5,*]{Peter G. Stockley}
\author[1,2,3,*]{Reidun Twarock}
\affil[1]{York Cross-disciplinary Centre for Systems Analysis, University of York, York YO10 5GE, UK}
\affil[2]{Department of Mathematics, University of York, York YO10 5DD, UK}
\affil[3]{Department of Biology, University of York, York YO10 5DD, UK}
\affil[4]{School of Science, Technology \& Health, York St John University, York YO31 7EX, UK}
\affil[5]{Astbury Centre for Structural Molecular Biology, University of Leeds, Leeds LS2 9JT, UK}

\affil[*]{corresponding authors: p.g.stockley@leeds.ac.uk (PGS) and rt507@york.ac.uk (RT)}

\affil[+]{these authors contributed equally to this work}

\keywords{}

\begin{abstract}
Defective interfering particles arise spontaneously during a viral infection as mutants lacking essential parts of the viral genome. Their ability to replicate in the presence of the wild-type (WT) virus (at the expense of viable viral particles) is mimicked and exploited by therapeutic interfering particles. We propose a strategy for the design of therapeutic interfering RNAs (tiRNAs) against positive-sense single-stranded RNA viruses that assemble via packaging signal-mediated assembly. These tiRNAs contain both an optimised version of the virus assembly manual that is encoded by multiple dispersed RNA packaging signals and a replication signal for viral polymerase,  but lack any protein coding information. We use an intracellular model for hepatitis C viral (HCV) infection that captures key aspects of the competition dynamics between tiRNAs and viral genomes for virally produced capsid protein and polymerase. We show that only a small increase in the assembly and replication efficiency of the tiRNAs compared with WT virus is required in order to achieve a treatment efficacy greater than 99\%. This demonstrates that the proposed tiRNA design could be a promising treatment option for RNA viral infections. 
\end{abstract}

\begin{document}

\maketitle
\section{Introduction}

Viruses are a major burden for public health and economy, yet our repertoire of antiviral options is still very limited. This is, in part, due to the high frequency with which viral genomes mutate and thus evade treatment. On the other hand, these mutations can sometimes lead to the production of defective viral genomes (DVGs) which are shed in defective interfering particles (DIPs). DVGs are spontaneously occurring mutants in a viral infection that lack essential genetic information, e.g. through deletion mutations, but are capable of replicating in the presence of, and indeed at the expense of, resources produced by viruses \cite{huang1970defective,poirier2017virus}. Many DVGs are well-known to have a replicative advantage over WT and to play a role in interference with WT virus \cite{li2021dengue,vignuzzi2019defective}, virus persistence \cite{manzoni2018defective} as well as specific \cite{dimmock2014defective} and unspecific immune activation \cite{manzoni2018defective,rand2021antiviral}. The exploitation of DIPs is a promising recent approach for therapy \cite{dimmock2014defective,rezelj2021defective,yao2021synthetic}. DIPs are selected and amplified for therapeutic use facilitated by advanced cloning techniques  \cite{dimmock2014defective,dimmock2012cloned, marriott2010defective,easton2011novel,mann2006interfering,noble2004interfering}, and have progressed to clinical stage \cite{ dimmock2012cloned,dimmock2014anti,dimmock2015method,dimmock2013cloned}.

{Von Magnus was the first to report the occurrence of DIPs, seen in influenza A virus populations passaged in embryonated chicken eggs \cite{von1954incomplete}. Subsequently, in a prolonged persistence of vesicular stomatitis virus infections, mediated by WT DIPs, mutants (called Sdi$^-$ mutants) were detected that are WT DIP resistant, demonstrating the importance of selection for DIP populations \cite{giachetti1988altered}. DIPs have been reported to cause oscillations in virus levels \cite{huang1973defective}. Zwart {\it et al.} \cite{zwart2013complex}  developed a simple mathematical model of baculovirus-DI dynamics which  qualitatively reproduced the oscillatory patterns seen in experimental data. DIPs engineered for therapy were designed to spread between individuals and autonomously target high-risk groups for HIV  and it has been argued that this method could decrease HIV/AIDS prevalence by 30-fold in 50 years \cite{metzger2011autonomous,notton2014case}. The impact of these transmissible antivirals are studied in intracellular, within-host, and epidemiological models \cite{rouzine2013design,rast2016conflicting}.}

\begin{figure}[H]
	\centering
	\includegraphics[width=0.72\linewidth]{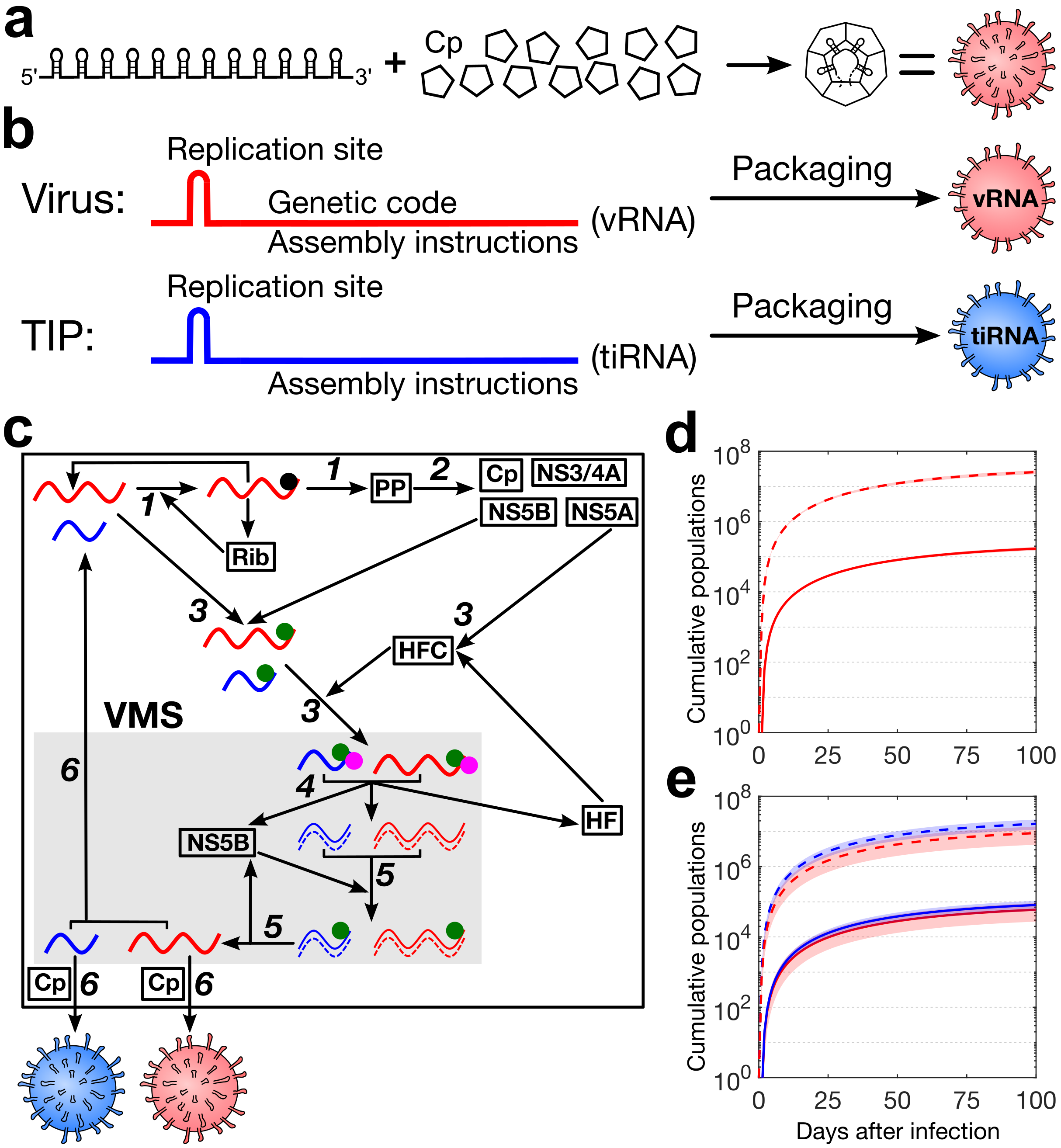}
	\caption{Assembly and intracellular dynamics of vRNAs and tiRNAs. (a) The PS-mediated assembly paradigm: Multiple sequence/structure motifs called packaging signals (PSs), that are dispersed throughout the viral genome, promote virion assembly via sequence specific interactions with coat protein (Cp). (b) vRNA and tiRNA in comparison: tiRNA is similar to vRNA but is devoid of any genetic message. (c) Schematic representation of the mathematical model for vRNA (red) and tiRNA (blue) in an HCV infection: In \emph{step 1} and \emph{2}; vRNA in the cytoplasm binds to free ribosomes to form a translation complex, which synthesizes the viral polyprotein (PP). The latter is cleaved, leading to the production of structural proteins such as core protein (Cp) and nonstructural proteins, including NS3/4A, NS5A, and NS5B. In \emph{step 3}; NS5B and NS5A bind to vRNA or tiRNA and host factor (HF), respectively. These two complexes are imported into the vesicular membranous structure (VMS). In \emph{step 4}; the imported RNAs form  double-strand RNAs (dsRNAs) and release NS5B and HF. In \emph{step 5}; dsRNAs again bind to the NS5B and synthesise new vRNAs and tiRNAs. In \emph{step 6}; these RNAs are either exported into the cytoplasm, or assembled into virions with 180 Cp and exported from the cell. (d) The time evolution of the HCV infection model shows the cumulative number of released virions (solid red line) and total vRNA (dashed red line), averaged over 250 simulations with the initial condition (+)RNA$^{\mbox{\scriptsize cyt}}$=1 and Cp=180. (e) The dynamics of virions and TIPs, where the solid red and blue lines indicate the released virions and TIPs, respectively, and the dashed red and blue lines the total vRNAs and tiRNAs, respectively, with the initial condition (+)RNA$^{\mbox{\scriptsize cyt}}$=1, (+)tiRNA$^{\mbox{\scriptsize cyt}}$=1 and Cp=360. The shaded areas highlight the regions of one standard deviation (std) from the mean. (d) and (e) are plotted using parameter values from Table \ref{param} with $b_a=1$ and $b_r=1$.}
	\label{fig1}
\end{figure}

{We propose here a novel strategy that exploits our discovery of packaging signal (PS)-mediated assembly in single-stranded RNA (ssRNA) viruses for the design of therapeutic interfering RNAs (tiRNAs) which are packaged into therapeutic interfering particles (TIPs) mimicking essential features of these DIPs. The genomes of many RNA viruses, including bacteriophages, plant viruses, and human pathogens present multiple dispersed sequence/structure motifs (PSs) that share a core recognition motif for the viral coat protein (Cp). Sequence variation around that core motif confers differential levels of affinity for Cp to the PSs, creating a hierarchy of affinities across the genome-embedded PS ensemble.  This enables PSs to recruit Cp to the growing capsid shell, collectively promoting virus assembly \cite{stockley2013new,dykeman2013packaging,stockley2013packaging,rolfsson2016direct, patel2017hbv, stewart2016identification, shakeel2017genomic} (Fig. \ref{fig1}a).}
\noindent
{Stochastic simulations of virus assembly revealed that the differences in affinity at distinct PS sites are important, because they bias the number of possible assembly pathways to a small number of dominant ones with very similar geometric properties, resulting in higher yields of fully assembled capsids compared with PS distributions lacking this variation. They also ensure specific encapsidation in the context of cellular competitor RNAs.   This means that PSs enable the virus to overcome the equivalent of Levinthal’s Paradox in protein folding. The latter refers to the complexity of protein folding, which would require an unrealistically long time to complete if it solely relied on a random exploration of all combinatorially possible pathways. Similarly, the assembly of a viral capsid would be too slow in the arms race against host defence mechanisms if it were exploring all geometrically possible pathways, rather than the subset favoured due to  interaction with the PSs \cite{dykeman2014solving}.} In some cases, such as in the Picornavirus human parechoviruses (HPeV), the PSs and their Cp binding sites have been characterised to atomistic detail \cite{shakeel2017genomic}. They have also been identified in hepatitis B virus \cite{patel2017hbv,fatehi2021intracellular}, a DNA virus that packages its genome into its capsid in the form of pre-genomic RNA. Our detailed mechanistic understanding of PS-mediated assembly has enabled us to optimise the assembly code in satellite tobacco necrosis virus (STNV), creating RNA mutants that outcompete viral genomes in a competition assay \cite{patel2017rewriting}. Many of the best-studied examples of DVGs are found in (-)ssRNA viruses, which are presently known not to assemble via multiple dispersed PSs. For viruses assembling via multiple dispersed PSs, indeed DIPs that are assembly competent and competitive with WT may not spontaneously arise through the standard mechanisms \cite{manzoni2018defective}; however, using our new mechanistic understanding they could be engineered. Here we propose to decouple the assembly code from the genetic message, and create synthetic assembly substrates containing only the PS-encoded virus assembly instructions allowing them to be efficiently encapsulated by viral Cp. To mimic the naturally occurring DIPs, we also include a recognition signal for viral replicase into our {\it de novo} designed tiRNAs, so that they are replicated by viral replicase (Fig. \ref{fig1}b). {Although some naturally occurring DIPs for influenza A \cite{dimmock2014defective} or Zika virus \cite{rezelj2021defective} have internal ribosome entry site (IRES), we deliberately exclude such IRESs here in order to give tiRNAs another competitive advantage over vRNAs by not spending time on translation. An added advantage is that the presence of IRESs, and consequently interactions between ribosomes and IRES containing tiRNAs, would trigger an immune response that could lead to the removal of the cell, unnecessarily increasing tissue damage.} 

Mathematical models of a viral infection can be used to assess the merits of novel antiviral strategies \cite{fatehi2021intracellular,Andino:2019}. The population dynamics of DIPs interacting with WT virus (called the `helper' or `standard' virus) can be complicated \cite{kirkwood1994cycles, bangham1990defective, thompson2010population,thompson2009multiple} with chaotic or predator-prey dynamics, but often the parasitic relationship between DIPs and helper virus results in the attenuation or clearance of the original infection. Thus, we use HCV (an ssRNA virus) as a model system and we develop an extension of an intracellular model of HCV presented by Aunins \emph{et al.} \cite{aunins2018intracellular} that now includes also the dynamic competition between tiRNAs and viral RNAs (vRNAs).

\section{Intracellular modelling of HCV infection and tiRNAs}

Recently Aunins \emph{et al.} presented a detailed, parameterised intracellular model for hepatitis C viral (HCV) infection based on experimental data \cite{aunins2018intracellular}. The model consists of 6 steps (Fig. \ref{fig1}c). \emph{Step 1}; positive-sense RNA strand in the cytoplasm [(+)RNA$^{\mbox{\scriptsize cyt}}$] binds to free ribosomes to form a translation complex (R:(+)RNA) at a rate $k_{tc}$, which synthesizes the viral polyprotein (PP) at a rate $k_{trans}$. \emph{Step 2}; the cleavage of PPs at a rate $k_{cleavage}$ leads to production of structural proteins, including core protein (Cp) and nonstructural proteins, including NS3/4A, NS5A, and NS5B$^{\mbox{\scriptsize cyt}}$.  \emph{Step 3}; NS5B$^{\mbox{\scriptsize cyt}}$ (polymerase) and NS5A bind to (+)RNA$^{\mbox{\scriptsize cyt}}$ and host factor (HF) at rates $k_{rp5b}$ and $k_{hfc}$ to form NS5B:(+)RNA and HFC complexes, respectively. These two complexes are imported into the vesicular membranous structure (VMS) at a rate $k_{rip}$ in a second-order reaction.  \emph{Step 4}; the imported RNA forms a double-strand RNA (dsRNA) and releases NS5B$^{\mbox{\scriptsize VMS}}$ and HF at a rate $k_{init}$.  \emph{Step 5}; dsRNA binds to the NS5B$^{\mbox{\scriptsize VMS}}$ at a rate $k_{rids}$ to synthesise (+)RNA$^{\mbox{\scriptsize VMS}}$ at a rate $k_{repl}$. \emph{Step 6}; (+)RNA$^{\mbox{\scriptsize VMS}}$ in the VMS are either exported into the cytoplasm at a rate $k_{outrp}$, or assembled into virions with 180 Cp and exported from the cell at a rate $k_{assembly}$. The model was fitted to experimental data and estimated parameters values (Table \ref{param})\cite{aunins2018intracellular}.

In this paper, we extend this model to include the dynamics of tiRNAs (Fig. \ref{fig1}c). Both the viral genome and tiRNA assemble via packaging signal (PS)-mediated assembly, i.e. both present PSs with affinity for viral Cp, ensuring efficient, specific genome packing. The PS distribution of the tiRNAs is optimised with respect to that of the virus, e.g. by stabilising key PSs in the distribution as in \cite{patel2017rewriting}, enabling them to potentially assemble more efficiently than the virus (Fig. \ref{fig1}a). We consider tiRNAs which also contain the terminal sequences necessary for recognition by viral polymerases \cite{dimmock2014defective,li2014sub,lin1993deletion}, so that they are replicated in the presence of the virus (Fig. \ref{fig1}b). As there is no protein coding requirement, and indeed no IRES, tiRNAs replication efficiency relative to the genome can be increased by shortening of the genome, or by tuning the nucleotide sequence \cite{li2014sub,lin1993deletion}. Thus, we model tiRNAs dynamics as follows: We assume that a positive-sense tiRNA in the cytoplasm [(+)tiRNA$^{\mbox{\scriptsize cyt}}$] binds to NS5B$^{\mbox{\scriptsize cyt}}$ and HFC at rates $k_{rp5b}$ and $k_{rip}$, respectively, to be imported into the VMS. Then, similar to viral RNAs (vRNAs), they produce double-strand tiRNA and (+)tiRNA$^{\mbox{\scriptsize VMS}}$ at rates $b_rk_{init}$ and $b_rk_{repl}$, respectively, where $b_r$ characterises the replication efficiency of the tiRNAs.  A tiRNA with $b_r=1$ would be transcribed at the same rate as vRNAs.  The newly formed (+)tiRNA$^{\mbox{\scriptsize VMS}}$ is then either exported into the cytoplasm at a rate $k_{outrp}$ or assembled into TIPs with 180 Cp and exported from the cell at a rate $b_ak_{assembly}$, where $b_a$ characterises the assembly efficiency of the tiRNAs (Fig. \ref{fig1}c). The assembly efficiency of a tiRNA can be tuned by alteration of the PS distribution, for instance, a competition experiment between STNV and a copy with an optimised PS distribution resulted in encapsidation in a ratio of $1:2$ to $1:3$ \cite{patel2017rewriting}, suggesting that $b_a \sim 2.5$ is experimentally achievable. The reactions of the model are provided in Methods.

Aunins \emph{et al.}\cite{aunins2018intracellular} used ordinary differential equations (ODE) to model the production of virions over a relatively short timescale (50 hours). {In this work we derive a continuous-time Markov chain (CTMC) model from the ODE model \cite{fatehi2018stochastic,yuan2011stochastic} and use the Gillespie algorithm as a discrete stochastic method for solving the CTMC model and plot stochastic trajectories \cite{gillespie1977exact}. The mean over the stochastic trajectories plotted using the Gillespie algorithm is in excellent agreement with the ODE dynamics (Supplementary Fig. \ref{FigS1}). However, using the Gillespie method will also enable the tracking of individual particles, which is of particular interest at low concentrations during the initial kinetic phase \cite{fatehi2021intracellular}. As the average life span of adult hepatocytes ranges from 200 to 300 days \cite{duncan2009stem} and since the half-life of HCV infected cells is estimated to be between 1.4 and 700 days \cite{dahari2009mathematical}, we ran our simulations for 100 days.}   

\section{Results}

\subsection{Single cell viral dynamics} 

Here we study the viral dynamics in the absence and presence of tiRNAs. The dynamics are computed as an average over {250} stochastic simulations of the reaction network using the Gillespie algorithm \cite{gillespie1977exact} implemented in Fortran and using parameter values from Table \ref{param} and initially, with $b_a=1$ and $b_r=1$. The multiplicity of infection (MOI) is set to one vRNA and one tiRNA, consistent with data for intranasal sprays for clonal influenza DIPs \cite{dimmock2014defective}. Figure \ref{fig1}d indicates the total number of released virions (solid red line) and vRNAs in the cell (dashed red line) in the absence of co-infecting tiRNAs. Figure \ref{fig1}e shows the effect on viral dynamics of introducing tiRNAs.  Co-infection with tiRNAs reduces the level of released virions by 70\%. {This reduction in the total number of released virions is called the treatment efficacy. It is here reached within 3 days, and remains within 2\% of this value thereafter.} The number of tiRNAs within the cell is comparable with the level of vRNAs in the tiRNA-free case. This shows that even without an advantage in replication or assembly, the lack of a protein-coding responsibility (and an IRES) enables tiRNAs to displace vRNAs as the most frequently packaged contents of new virus-like particles (VLPs).   

The impact of the tiRNAs relies on two characteristic features: their relative replication ($b_r$) and assembly ($b_a$) efficiency compared with helper virus. We therefore investigate the impact of these two descriptors on the infection dynamics.

\begin{figure}[H]
	\centering
	\includegraphics[width=0.73\linewidth]{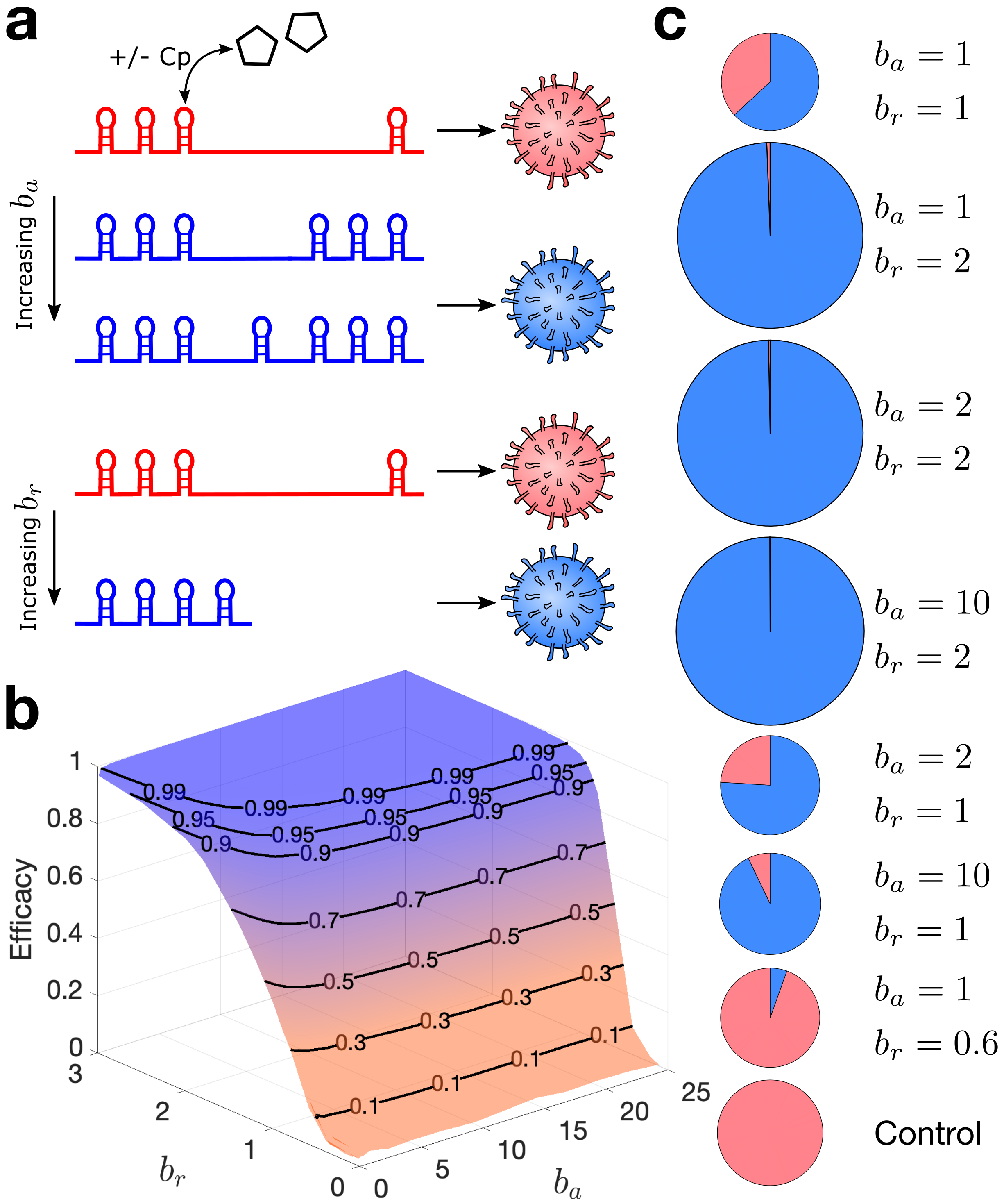}
	\caption{The impact of replication and assembly efficiencies on the treatment efficacy and cumulative number of released virus-like particles. (a) TIP design is defined by two parameters: assembly efficacy $b_a$, which can be changed by addition of PSs in the tiRNAs (blue) compared with vRNA (red), and replication efficacy $b_r$, which can be improved by shortening of tiRNA with respect to vRNA. (b) The efficacy of tiRNAs (shown as the fraction indicating reduction in the number of released infectious virions) as a function of $b_a$ and $b_r$. (c) Pie charts for the cumulative number of released virions and TIPs after 100 days post infection. Red and blue indicate virions and TIPs, respectively. The tiRNA-free control is shown for comparison. The area of each graph is proportional to the total number of released particles (virion+TIP) with respect to the control. For $b_r\leq 1$ the total number of released particles is less than the control while for $b_r=2$ the total number of released particles is 3 times of the control. (b) and (c) are plotted by averaging over 250 simulations with the initial condition (+)RNA$^{\mbox{\scriptsize cyt}}$=1, (+)tiRNA$^{\mbox{\scriptsize cyt}}$=1 and Cp=360, using parameter values from Table \ref{param}.}
	\label{fig2}
\end{figure}

\subsection{The effect of replication and assembly efficiencies} 

tiRNAs assembly and replication efficiency can be increased by adding more PSs into the genome, stabilising PSs or increasing their binding affinity, and shortening of the genome, respectively (Fig. \ref{fig2}a). Though many DVGs are well-known to have a replicative advantage over WT {probably due to their shorter genomes \cite{manzoni2018defective,rezelj2018defective,rezelj2021defective}, the length of the genome is not the only factor that determines the replication efficiency ($b_r$) \cite{thomson1998genomic,lin1993deletion}}. Thus, we also consider the cases where $b_r<1$, i.e., where tiRNAs replicate less efficiently than WT virus. Figure \ref{fig2}b indicates for $b_r<0.7$ the efficacy of tiRNAs is below 50\%, even for high values of $b_a$ (assembly efficiency). This illustrates the importance of the replication process in the viral life cycle, as tiRNAs must be at least as efficient as WT virus at replication in order to be a viable treatment option. For $b_r\leq 1$, i.e. if tiRNAs are not more efficient at replication than virus, the total number of released particles is lower than the tiRNA-free control, while for $b_r>1$ the total number of released particles increases, however this is overwhelmingly dominated by TIPs (Fig. \ref{fig2}c). This increase in the number of TIPs will increase the level of antigen and could have consequences for the immune response. However, increasing of $b_a$  alone does not lead to an increase in the level of total released particles (Fig. \ref{fig2}c). {As the length of the genome is not the only factor impacting $b_r$, we have done our analysis for varying values of $b_r$. We note that there are some naturally occurring DVGs that are 30\% shorter than the WT genome. If  $b_r$ were related to length linearly, this would give $b_r=1.428$ \cite{rezelj2021defective}. Thus, considering values up to $b_r=2$ for the  {\it de novo} designed tiRNAs seems achievable}. Interestingly,  the benefits of the treatment have a saturation point; increasing $b_a>10$ and increasing $b_r>2$ does not have a significant impact on the efficacy of the treatment and total number of released particles (Fig. \ref{fig2}b and c).

Even for $b_a=b_r=1$, i.e. for equal replication and assembly efficiency as the virus, the number of released TIPs is higher than that of infectious virions. This is because tiRNAs have a competitive advantage over virus as they are depleting resources generated only by the virus, using virally generated polymerase for replication and Cp for assembly. In particular, during the time that vRNA is bound to ribosome, tiRNA is free to bind to NS5B and replicate at the expense of the virus, resulting in the inherent asymmetry between virus and TIP. 

\subsection{The effect of higher multiplicities of infection (MOIs)}

The above results have been obtained in the equitable case of an MOI of 1:1.  Experimental work on comparable systems have reported DIP MOIs in the range $1-100$ \cite{dimmock2014defective, thompson2010population,abrahao2018tailed}, therefore we should consider cases with unequal starting proportions of vRNA and tiRNA. We next determine the release kinetics for higher MOIs, setting $b_a=b_r=1$ in order to isolate the effect of the MOI. 

\begin{figure}[H]
	\centering
	\includegraphics[width=0.62\linewidth]{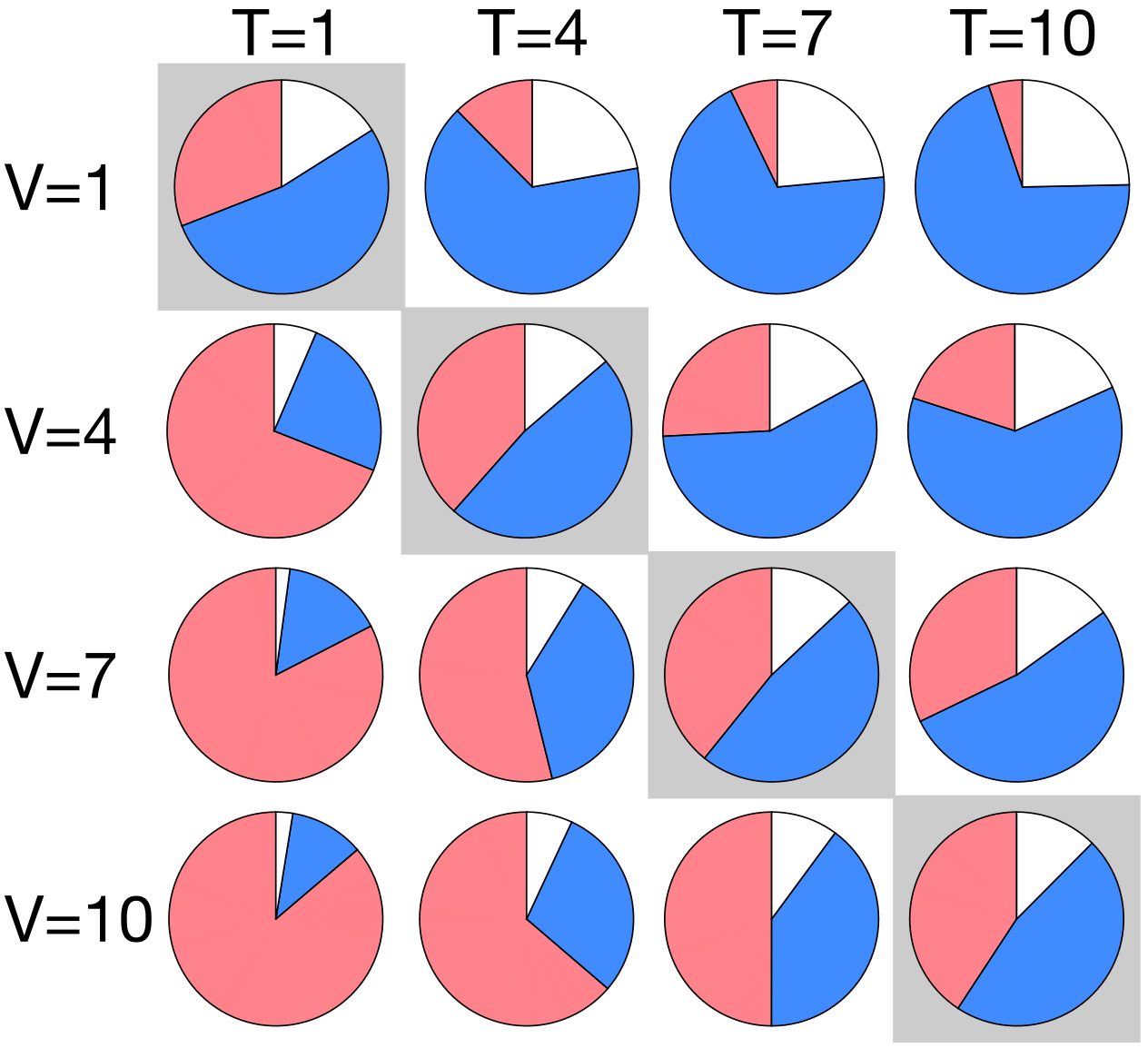}
	\caption{Increasing the MOI of tiRNA (T) increases the cumulative number of released TIPs and reduces the total number of released particles (virion+TIP) compared with the control. Pie charts for the cumulative number of released virions and TIPs after 100 days post infection. Red and blue indicate the number of virions and TIPs, respectively, while white shows the difference between the total number of released virions in the control (tiRNA-free) case with the number of released particles (virion+TIP) in the presence of treatment. This figure is plotted by averaging over 250 simulations using parameter values from Table \ref{param} with $b_a=1$ and $b_r=1$. The initial condition for the MOIs of V=$m$, T=$n$ is (+)RNA$^{\mbox{\scriptsize cyt}}$=$m$, (+)tiRNA$^{\mbox{\scriptsize cyt}}$=$n$, and Cp=$180(m+n)$.}
	\label{fig3}
\end{figure}

For MOIs of tiRNA (T) higher than the MOI of vRNA (V), the number of released virions decreases while increasing the ratio of TIPs/Virions (Fig. \ref{fig3}), with TIPs swiftly dominating the population. However, for MOIs of vRNA larger than the MOI of tiRNA, a much smaller effect occurs (Fig. \ref{fig3}) and virions outnumber the TIPs, demonstrating that the relative ratio of virus and TIPs is important for the outcome of the treatment. From Fig. \ref{fig3} we can see that when the MOI of tiRNA and vRNA is equal V=T=1 the released particles are dominated by TIPs in a roughly 1.5:1 ratio with the virions.  This suggests that in a population of infected cells, the subsequent infections would also be seeded by MOIs with more tiRNAs than vRNAs, moving rightward in Fig. \ref{fig3}, leading to a population of released particles dominated by non-infectious TIPs, potentially causing the elimination of the wider infection.  A within-host model of the HCV in the presence of tiRNAs is required to fully examine this potential.    

\subsection{The effect of treatment starting time}

\begin{figure}[H]
	\centering
	\includegraphics[width=0.7\linewidth]{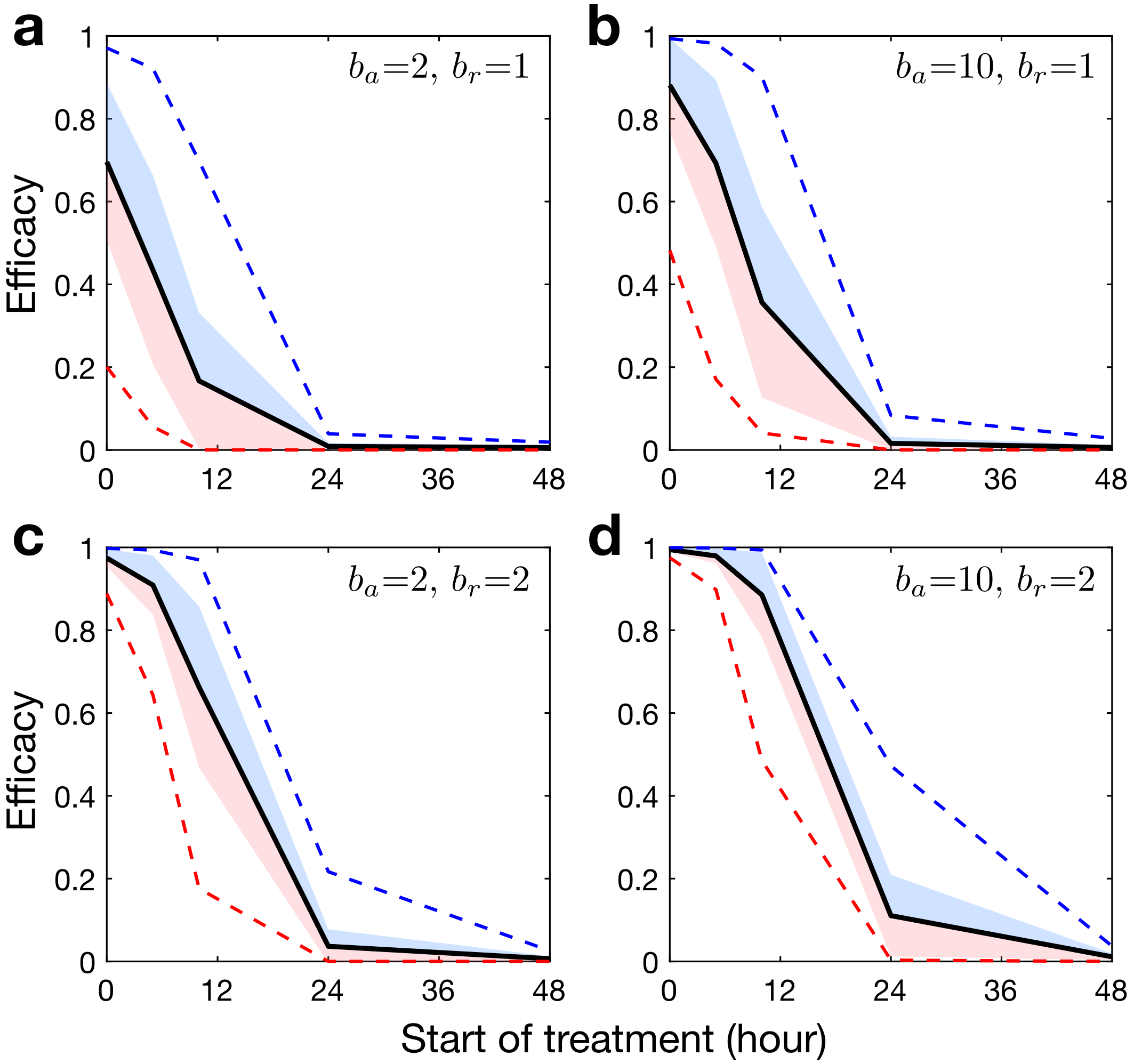}
	\caption{Starting treatment 24 hours after cell infection has no significant impact even for high replication and assembly efficiencies and MOIs. The black curves indicate the average of efficacy over varying MOI of vRNAs and tiRNAs from 1 to 10 ($1\leq$V$\leq 10$, $1\leq$T$\leq 10$). Blue and red shaded areas show the regions of mean+std and mean-std, respectively. Blue and red dashed lines indicates the maximum and minimum efficacy of treatment over various MOIs, respectively.}
	\label{fig4}
\end{figure}

The results presented above are based on the assumption that both the vRNAs and the tiRNAs begin the infection at the same time, however this would not necessarily be the case \emph{in vivo}. Figure \ref{fig4} shows that if the MOI of tiRNA is larger than that of vRNA (blue shaded area) the efficacy is higher than the average efficacy (solid black line). On the other hand, while if the MOI of vRNA is larger than that of tiRNA (red shaded area) the efficacy is lower than the average efficacy. If each cell that is infected by a vRNA already harbours at least one tiRNA (start of treatment = 0 hour), we get the highest treatment efficacy (Fig. \ref{fig4}). {However, if treatment was started (i.e. the tiRNAs introduced) after 24 hours post infection, this treatment option has no significant effect on the outcome of the infection even if $b_a$, $b_r$ are high and the MOI of tiRNAs is greater than that of the vRNAs. This is because there are more than 1,000 vRNAs in the cell, so that tiRNAs have little chance to overtake the viral life cycle (Fig. \ref{fig4}). This suggests that tiRNAs can be highly effective when used as a prophylactic antiviral treatment, an approach that has recently gained wider attention \cite{dimmock2014defective,czuppon2021success}}. Recent experiments for influenza DIPs have established that prophylactic intranasal treatment can achieve delivery of around one DIP per susceptible cell, where they can stay present for several weeks \cite{dimmock2014defective}. Although evaluation of the full impact of TIPs as a treatment option during a chronic infection needs to be studied in the context of a within-host model, our model provides the foundation for studying such aspects by coupling of the intracellular model presented here with an intercellular model in a multiscale approach \cite{quintela2018new,fatehi2021comparing}.

\section{Discussion}

In this paper we have exploited recent insights into PS-mediated assembly in order to propose a novel design for TIPs that combines the replicative advantages of existing DIP/TIP strategies over viruses \cite{rast2016conflicting,dimmock2014defective,marriott2010defective} with the benefits of PS-mediated assembly. This novel design for TIPs opens up unprecedented therapeutic potential to interfere with viral replication and assembly, and misdirect viral resources. We have demonstrated the benefits of this combined strategy through {stochastic simulations}. 

{The tiRNA designs proposed here are alternatives to other therapeutic strategies exploiting PS-mediated assembly, such as small molecular weight compounds targeting the PS:RNA contacts. PS motifs are shared by viruses from the same family (e.g. in picornaviruses\cite{shakeel2017genomic,chandlerbostock2020}). Some viruses have been repurposed by nature for vital functions in the host organism. For instance,  captured retroviruses and retrotransposons can be expressed in different tissues and at different points during life cycles, such as in neuronal functioning \cite{pastuzyn2018neuronal,ashley2018retrovirus} or  placentation \cite{grow2015intrinsic,cornelis2017endogenous,cornelis2015retroviral,denner2017function}. Any PS-targeting drugs interfering with virus assembly in related viruses would therefore have to be monitored carefully with regards to their effect on other vital functions. tiRNAs, on the other hand, would not pose that risk.}

However, they have the same benefits as PS-targeting drugs from the point of view of viral escape. This is because in both cases, viral escape would require mutation of the full set (or a significant subset) of the PSs, which would result in a significant intermittent fitness loss due to the multiple dispersed nature of the PSs \cite{bingham2017rna}, making such a transition highly unlikely. The low propensity for therapy resistance is shared by other DIPs/TIPs, where resistant strains have been shown to be selected against at both individual and population level \cite{rast2016conflicting}. However, our tiRNA design has another  advantage: Being stripped of all genetic information, they do not pose the risk of recombination. This is in contrast with DIPs that arise via deletion of a portion of the viral sequence, such as one genomic segment in the case of the multi-segmented influenza virus \cite{dimmock2014defective}, and that retain the coding capacity for some of the gene products.

{As we show here, tiRNAs show promise as an effective treatment at the intracellular level, especially as a prophylactic treatment. Our intracellular model shows that in this case the tiRNA only needs to be as effective at packaging and replicating as vRNA to dominate the population of particles released from the infected cell.} {DVGs arising naturally in viruses that assemble via multiple dispersed packaging signals may have only a subset of the PS sites and therefore not be optimal at assembly; the possibility to \emph{de novo}-engineer tiRNAs that not only are assembly-competent but are more assembly-efficient than WT as well as having a replicative advantage therefore constitutes a step-change for antiviral approaches for this group of viruses.} {As TIPs are indistinguishable from WT virus on the particle exterior, they can elicit an immune response (increasing antigen levels), thus priming the immune system against a subsequent infection. Multiscale within-host models of the immune response will be required to study the impact of this within infected patients, and analyse the impact of TIPs on both acute and chronic infections. Furthermore,  models of between-host infection dynamics will reveal the consequences of any TIP transmission between hosts, and thus potentially reveal  additional benefits to the host population arising from this. Such models would also enable evaluation of risks associated with the use of TIPs \cite{kirkwood1994cycles,thompson2010population,rast2016conflicting}.}

\section{Methods}

\subsection{Reaction kinetics of the viral life cycle}
\label{reactions}
The first step is the production of polyprotein (PP) using the host cell ribosomes (Rib) from the (+)RNA in cytoplasm [(+)RNA$^{\mbox{\scriptsize cyt}}$]:
\begin{align*}
	&\mbox{(+)RNA}^{\mbox{\scriptsize cyt}}+\mbox{Rib}\xrightarrow{k_{tc}}\mbox{R:(+)RNA},\\
	&\mbox{R:(+)RNA}\xrightarrow{k_{trans}}\mbox{PP}+\mbox{Rib}+\mbox{(+)RNA}^{\mbox{\scriptsize cyt}}.
\end{align*}
\noindent
Then, the viral proteins are cleaved in a single reaction from the PP:
\begin{align*}
	\mbox{PP}\xrightarrow{k_{cleavage}}\mbox{Cp}+\mbox{NS3/4A}+\mbox{NS5A}+\mbox{NS5B}^{\mbox{\scriptsize cyt}}.
\end{align*}
\noindent
In the next step, NS5A and NS5B$^{\mbox{\scriptsize cyt}}$ bind to human factor (HF) and (+)RNA$^{\mbox{\scriptsize cyt}}$ or (+)tiRNA$^{\mbox{\scriptsize cyt}}$, respectively, and these two complexes are imported into the VMS in a second-order reaction:
\begin{align*}
	&\mbox{(+)RNA}^{\mbox{\scriptsize cyt}}+\mbox{NS5B}^{\mbox{\scriptsize cyt}}\xrightarrow{k_{rp5b}}\mbox{NS5B:(+)RNA},\\
	&\mbox{(+)tiRNA}^{\mbox{\scriptsize cyt}}+\mbox{NS5B}^{\mbox{\scriptsize cyt}}\xrightarrow{k_{rp5b}}\mbox{NS5B:(+)tiRNA},\\
	&\mbox{HF}+\mbox{NS5A}\xrightarrow{k_{hfc}}\mbox{HFC},\\
	&\mbox{NS5B:(+)RNA}+\mbox{HFC}\xrightarrow{k_{rip}}\mbox{NS5B:(+)RNA:HFC},\\
	&\mbox{NS5B:(+)tiRNA}+\mbox{HFC}\xrightarrow{k_{rip}}\mbox{NS5B:(+)tiRNA:HFC}.
\end{align*}
\noindent
The imported RNA (tiRNA) form a dsRNA (dstiRNA) and releases NS5B and HF:
\begin{align*}
	&\mbox{NS5B:(+)RNA:HFC}\xrightarrow{k_{init}}\mbox{NS5B}^{\mbox{\scriptsize VMS}}+\mbox{dsRNA}+\mbox{HF},\\
	&\mbox{NS5B:(+)tiRNA:HFC}\xrightarrow{b_rk_{init}}\mbox{NS5B}^{\mbox{\scriptsize VMS}}+\mbox{dstiRNA}+\mbox{HF}.
\end{align*}
\noindent
Once again the dsRNA (dstiRNA) and NS5B$^{\mbox{\scriptsize VMS}}$ form a complex required to synthesise new (+)RNA$^{\mbox{\scriptsize VMS}}$ ((+)tiRNA$^{\mbox{\scriptsize VMS}}$):
\begin{align*}
	&\mbox{dsRNA}+\mbox{NS5B}^{\mbox{\scriptsize VMS}}\xrightarrow{k_{rids}}\mbox{NS5B:dsRNA},\\
	&\mbox{dstiRNA}+\mbox{NS5B}^{\mbox{\scriptsize VMS}}\xrightarrow{k_{rids}}\mbox{NS5B:dstiRNA},\\
	&\mbox{NS5B:dsRNA}\xrightarrow{k_{repl}}\mbox{dsRNA}+\mbox{NS5B}^{\mbox{\scriptsize VMS}}+\mbox{(+)RNA}^{\mbox{\scriptsize VMS}},\\
	&\mbox{NS5B:dstiRNA}\xrightarrow{b_rk_{repl}}\mbox{dstiRNA}+\mbox{NS5B}^{\mbox{\scriptsize VMS}}+\mbox{(+)tiRNA}^{\mbox{\scriptsize VMS}}.
\end{align*}
\noindent
(+)RNA$^{\mbox{\scriptsize VMS}}$ ((+)tiRNA$^{\mbox{\scriptsize VMS}}$) either exported into the cytoplasm or assembled into new virions (TIPs) with Cp:
\begin{align*}
	&\mbox{(+)RNA}^{\mbox{\scriptsize VMS}}\xrightarrow{k_{outrp}}\mbox{(+)RNA}^{\mbox{\scriptsize cyt}},\\
	&\mbox{(+)tiRNA}^{\mbox{\scriptsize VMS}}\xrightarrow{k_{outrp}}\mbox{(+)tiRNA}^{\mbox{\scriptsize cyt}},\\
	&\mbox{(+)RNA}^{\mbox{\scriptsize VMS}}+180\mbox{Cp}\xrightarrow{k_{assembly}}\mbox{virion},\\
	&\mbox{(+)tiRNA}^{\mbox{\scriptsize VMS}}+180\mbox{Cp}\xrightarrow{b_ak_{assembly}}\mbox{TIP}.
\end{align*}
\noindent
Finally, we allow natural decay of vRNAs, tiRNAs, proteins and VMS via the reactions:
\begin{equation*}
\begin{aligned}[c]
	&\mbox{(+)RNA}^{\mbox{\scriptsize cyt}}\xrightarrow{k_{degrp}}0,\\
	&\mbox{NS5B:(+)RNA:HFC}\xrightarrow{k_{degvms}}0,\\
	&\mbox{dsRNA}\xrightarrow{k_{degvms}}0,\\
	&\mbox{(+)RNA}^{\mbox{\scriptsize VMS}}\xrightarrow{k_{degvms}}0,\\
	&\mbox{NS5B:dsRNA}\xrightarrow{k_{degvms}}0,\\
	&\mbox{NS3/4A}\xrightarrow{k_{degns}}0,\\
	&\mbox{NS5B}^{\mbox{\scriptsize cyt}}\xrightarrow{k_{degns}}0,\\
	&\mbox{Cp}\xrightarrow{k_{degs}}0.
\end{aligned}
\qquad\qquad
\begin{aligned}[c]
   &\mbox{(+)tiRNA}^{\mbox{\scriptsize cyt}}\xrightarrow{k_{degrp}}0,\\
   &\mbox{NS5B:(+)tiRNA:HFC}\xrightarrow{k_{degvms}}0,\\
   &\mbox{dstiRNA}\xrightarrow{k_{degvms}}0,\\
   &\mbox{(+)tiRNA}^{\mbox{\scriptsize VMS}}\xrightarrow{k_{degvms}}0,\\
   &\mbox{NS5B:dstiRNA}\xrightarrow{k_{degvms}}0,\\
   &\mbox{NS5A}\xrightarrow{k_{degns}}0,\\
   &\mbox{NS5B}^{\mbox{\scriptsize VMS}}\xrightarrow{k_{degvms}}0,\\
   &
\end{aligned}
\end{equation*}

\subsection{Parameter values}
\label{param section}
We use parameter values that have been reported in Aunins {\it et al.} \cite{aunins2018intracellular} and are presented in Table \ref{param}.

\begin{table}[H]
\centering
\caption{Table of parameter values}
\label{param}
\begin{tabular}{l|l|l|l}
\hline\hline
Parameter      & Value                                                           & Parameter    & Value                           \\\hline
$k_{tc}$       & 1 molecule$^{-1}$ h$^{-1}$                                      & $k_{trans}$  & 180 h$^{-1}$                    \\
$k_{cleavage}$ & 9 h$^{-1}$                                                      & $k_{hfc}$    & 0.0008 molecule$^{-1}$ h$^{-1}$ \\
$k_{rp5b}$     & 0.1 molecule$^{-1}$ h$^{-1}$                                    & $k_{rip}$    & 0.6 molecule$^{-1}$ h$^{-1}$    \\
$k_{init}$     & 1.12 h$^{-1}$                                                   & $k_{rids}$   & 10 molecule$^{-1}$ h$^{-1}$     \\
$k_{repl}$     & 1.12 h$^{-1}$                                                   & $k_{outrp}$  & 0.307 h$^{-1}$                  \\
$k_{assembly}$ & $1.2\times 10^{-7}$ molecule$^{-1}$ h$^{-1}$                    & $k_{degrp}$  & 0.26 h$^{-1}$                   \\
$k_{degns}$    & 0.11 h$^{-1}$                                                   & $k_{degvms}$ & 0.001 h$^{-1}$                  \\
Rib            & 5000 molecules                                                  & HF           & 30 molecules                    \\\hline
$k_{degs}$     & \multicolumn{3}{l}{0.61 h$^{-1}$ from 0 to 21 h, and 0.1 h$^{-1}$ from 21 h onward}\\\hline                            
\end{tabular}
\end{table}

\section*{Data availability}

All data generated or analysed during this study are included in this published article (and its Supplementary Information file).

\section*{Acknowledgements}

PGS and RT thank the Wellcome Trust for financial support through the Joint Investigator Award (110145 \& 110146), which also provided funding for P-PD and FF. Moreover, RT thanks the EPSRC for an Established Career Fellowship (EP/R023204/1) and the Royal Society for a Royal Society Wolfson Fellowship (RSWF/R1/180009). 

\section*{Author contributions statement}

R.T., P.-P.D., and P.G.S. conceived the goals of the study. F.F. and R.J.B. conducted the analysis. F.F., R.J.B., R.T., and P.-P.D. analysed the results. All authors reviewed the manuscript.

\section*{Competing interests}

The authors declare no conflict of interest.

\section*{Accession codes}

All codes are available in the authors' GitHub page:\\ \href{https://github.com/MathematicalComputationalVirology/HCVIntracellularModelling}{github.com/MathematicalComputationalVirology/HCVIntracellularModelling}


\begin{thebibliography}{62}
\providecommand{\natexlab}[1]{#1}
\providecommand{\url}[1]{\texttt{#1}}
\expandafter\ifx\csname urlstyle\endcsname\relax
  \providecommand{\doi}[1]{doi: #1}\else
  \providecommand{\doi}{doi: \begingroup \urlstyle{rm}\Url}\fi

\bibitem[Abrah{\~a}o et~al.(2018)Abrah{\~a}o, Silva, Silva, Khalil, Rodrigues,
  Arantes, Assis, Boratto, Andrade, Kroon, et~al.]{abrahao2018tailed}
J.~Abrah{\~a}o, L.~Silva, L.~S. Silva, J.~Y.~B. Khalil, R.~Rodrigues,
  T.~Arantes, F.~Assis, P.~Boratto, M.~Andrade, E.~G. Kroon, et~al.
\newblock Tailed giant {Tupanvirus} possesses the most complete translational
  apparatus of the known virosphere.
\newblock \emph{Nat. Commun.}, 9\penalty0 (1):\penalty0 749, 2018.

\bibitem[Ashley et~al.(2018)Ashley, Cordy, Lucia, Fradkin, Budnik, and
  Thomson]{ashley2018retrovirus}
J.~Ashley, B.~Cordy, D.~Lucia, L.~G. Fradkin, V.~Budnik, and T.~Thomson.
\newblock Retrovirus-like {Gag} protein {Arc1} binds {RNA} and traffics across
  synaptic boutons.
\newblock \emph{Cell}, 172\penalty0 (1):\penalty0 262--274, 2018.

\bibitem[Aunins et~al.(2018)Aunins, Marsh, Subramanya, Uprichard, Perelson, and
  Chatterjee]{aunins2018intracellular}
T.~R. Aunins, K.~A. Marsh, G.~Subramanya, S.~L. Uprichard, A.~S. Perelson, and
  A.~Chatterjee.
\newblock Intracellular hepatitis {C} virus modeling predicts infection
  dynamics and viral protein mechanisms.
\newblock \emph{J. Virol.}, 92\penalty0 (11), 2018.

\bibitem[Bangham and Kirkwood(1990)]{bangham1990defective}
C.~R.~M. Bangham and T.~B.~L. Kirkwood.
\newblock Defective interfering particles: effects in modulating virus growth
  and persistence.
\newblock \emph{Virology}, 179\penalty0 (2):\penalty0 821--826, 1990.

\bibitem[Bingham et~al.(2017)Bingham, Dykeman, and Twarock]{bingham2017rna}
R.~J. Bingham, E.~C. Dykeman, and R.~Twarock.
\newblock {RNA} virus evolution via a quasispecies-based model reveals a drug
  target with a high barrier to resistance.
\newblock \emph{Viruses}, 9\penalty0 (11):\penalty0 347, 2017.

\bibitem[Chandler-Bostock et~al.(2020)Chandler-Bostock, Mata, Bingham, Dykeman,
  Meng, Tuthill, Rowlands, Ranson, Twarock, and Stockley]{chandlerbostock2020}
R.~Chandler-Bostock, C.~P. Mata, R.~J. Bingham, E.~C. Dykeman, B.~Meng, T.~J.
  Tuthill, D.~J. Rowlands, N.~A. Ranson, R.~Twarock, and P.~G. Stockley.
\newblock Assembly of infectious enteroviruses depends on multiple, conserved
  genomic {RNA}-coat protein contacts.
\newblock \emph{PLoS Pathog.}, 16\penalty0 (12):\penalty0 e1009146, 2020.

\bibitem[Cornelis et~al.(2015)Cornelis, Vernochet, Carradec, Souquere, Mulot,
  Catzeflis, Nilsson, Menzies, Renfree, Pierron,
  et~al.]{cornelis2015retroviral}
G.~Cornelis, C.~Vernochet, Q.~Carradec, S.~Souquere, B.~Mulot, F.~Catzeflis,
  M.~A. Nilsson, B.~R. Menzies, M.~B. Renfree, G.~Pierron, et~al.
\newblock Retroviral envelope gene captures and syncytin exaptation for
  placentation in marsupials.
\newblock \emph{Proc. Natl. Acad. Sci.}, 112\penalty0 (5):\penalty0 E487--E496,
  2015.

\bibitem[Cornelis et~al.(2017)Cornelis, Funk, Vernochet, Leal, Tarazona,
  Meurice, Heidmann, Dupressoir, Miralles, Ramirez-Pinilla,
  et~al.]{cornelis2017endogenous}
G.~Cornelis, M.~Funk, C.~Vernochet, F.~Leal, O.~A. Tarazona, G.~Meurice,
  O.~Heidmann, A.~Dupressoir, A.~Miralles, M.~P. Ramirez-Pinilla, et~al.
\newblock An endogenous retroviral envelope syncytin and its cognate receptor
  identified in the viviparous placental {Mabuya} lizard.
\newblock \emph{Proc. Natl. Acad. Sci.}, 114\penalty0 (51):\penalty0
  E10991--E11000, 2017.

\bibitem[Czuppon et~al.(2021)Czuppon, D{\'e}barre, Gon{\c{c}}alves, Tenaillon,
  Perelson, Guedj, and Blanquart]{czuppon2021success}
P.~Czuppon, F.~D{\'e}barre, A.~Gon{\c{c}}alves, O.~Tenaillon, A.~S. Perelson,
  J.~Guedj, and F.~Blanquart.
\newblock Success of prophylactic antiviral therapy for {SARS-CoV}-2: Predicted
  critical efficacies and impact of different drug-specific mechanisms of
  action.
\newblock \emph{PLoS Comput. Biol.}, 17\penalty0 (3):\penalty0 e1008752, 2021.

\bibitem[Dahari et~al.(2009)Dahari, Layden-Almer, Kallwitz, Ribeiro, Cotler,
  Layden, and Perelson]{dahari2009mathematical}
H.~Dahari, J.~E. Layden-Almer, E.~Kallwitz, R.~M. Ribeiro, S.~J. Cotler, T.~J.
  Layden, and A.~S. Perelson.
\newblock A mathematical model of hepatitis {C} virus dynamics in patients with
  high baseline viral loads or advanced liver disease.
\newblock \emph{Gastroenterology}, 136\penalty0 (4):\penalty0 1402--1409, 2009.

\bibitem[Denner(2017)]{denner2017function}
J.~Denner.
\newblock Function of a retroviral envelope protein in the placenta of a
  viviparous lizard.
\newblock \emph{Proc. Natl. Acad. Sci.}, 114\penalty0 (51):\penalty0
  13315--13317, 2017.

\bibitem[Dimmock(2013)]{dimmock2013cloned}
N.~Dimmock.
\newblock Cloned defective interfering influenza {A} virus, May~7 2013.
\newblock {US} Patent 8,435,508.

\bibitem[Dimmock(2015)]{dimmock2015method}
N.~Dimmock.
\newblock Method of preventing or treating influenza {A} viral infection using
  cloned {DI} influenza {A} viral particles, July~28 2015.
\newblock {US} Patent 9,089,516.

\bibitem[Dimmock and Easton(2014{\natexlab{a}})]{dimmock2014anti}
N.~Dimmock and A.~Easton.
\newblock Anti-viral protection with viruses containing defective genome
  segments, Apr.~8 2014{\natexlab{a}}.
\newblock {US} Patent 8,691,215.

\bibitem[Dimmock and Easton(2014{\natexlab{b}})]{dimmock2014defective}
N.~J. Dimmock and A.~J. Easton.
\newblock Defective interfering influenza virus {RNAs}: time to reevaluate
  their clinical potential as broad-spectrum antivirals?
\newblock \emph{J. Virol.}, 88\penalty0 (10):\penalty0 5217--5227,
  2014{\natexlab{b}}.

\bibitem[Dimmock et~al.(2012)Dimmock, Dove, Scott, Meng, Taylor, Cheung,
  Hallis, Marriott, Carroll, and Easton]{dimmock2012cloned}
N.~J. Dimmock, B.~K. Dove, P.~D. Scott, B.~Meng, I.~Taylor, L.~Cheung,
  B.~Hallis, A.~C. Marriott, M.~W. Carroll, and A.~J. Easton.
\newblock Cloned defective interfering influenza virus protects ferrets from
  pandemic 2009 influenza {A} virus and allows protective immunity to be
  established.
\newblock \emph{PLoS ONE}, 7\penalty0 (12):\penalty0 e49394, 2012.

\bibitem[Duncan et~al.(2009)Duncan, Dorrell, and Grompe]{duncan2009stem}
A.~W. Duncan, C.~Dorrell, and M.~Grompe.
\newblock Stem cells and liver regeneration.
\newblock \emph{Gastroenterology}, 137\penalty0 (2):\penalty0 466--481, 2009.

\bibitem[Dykeman et~al.(2013)Dykeman, Stockley, and
  Twarock]{dykeman2013packaging}
E.~C. Dykeman, P.~G. Stockley, and R.~Twarock.
\newblock Packaging signals in two single-stranded {RNA} viruses imply a
  conserved assembly mechanism and geometry of the packaged genome.
\newblock \emph{J. Mol. Biol.}, 425\penalty0 (17):\penalty0 3235--3249, 2013.

\bibitem[Dykeman et~al.(2014)Dykeman, Stockley, and
  Twarock]{dykeman2014solving}
E.~C. Dykeman, P.~G. Stockley, and R.~Twarock.
\newblock Solving a {Levinthal}'s paradox for virus assembly identifies a
  unique antiviral strategy.
\newblock \emph{Proc. Natl. Acad. Sci.}, 111\penalty0 (14):\penalty0
  5361--5366, 2014.

\bibitem[Easton et~al.(2011)Easton, Scott, Edworthy, Meng, Marriott, and
  Dimmock]{easton2011novel}
A.~J. Easton, P.~D. Scott, N.~L. Edworthy, B.~Meng, A.~C. Marriott, and N.~J.
  Dimmock.
\newblock A novel broad-spectrum treatment for respiratory virus infections:
  influenza-based defective interfering virus provides protection against
  pneumovirus infection in vivo.
\newblock \emph{Vaccine}, 29\penalty0 (15):\penalty0 2777--2784, 2011.

\bibitem[Fatehi et~al.(2018)Fatehi, Kyrychko, Ross, Kyrychko, and
  Blyuss]{fatehi2018stochastic}
F.~Fatehi, S.~N. Kyrychko, A.~Ross, Y.~N. Kyrychko, and K.~B. Blyuss.
\newblock Stochastic effects in autoimmune dynamics.
\newblock \emph{Front. Physiol.}, 9:\penalty0 45, 2018.

\bibitem[Fatehi et~al.(2021{\natexlab{a}})Fatehi, Bingham, Dykeman, Patel,
  Stockley, and Twarock]{fatehi2021intracellular}
F.~Fatehi, R.~J. Bingham, E.~C. Dykeman, N.~Patel, P.~G. Stockley, and
  R.~Twarock.
\newblock An intracellular model of hepatitis {B} viral infection: An in silico
  platform for comparing therapeutic strategies.
\newblock \emph{Viruses}, 13\penalty0 (1):\penalty0 11, 2021{\natexlab{a}}.

\bibitem[Fatehi et~al.(2021{\natexlab{b}})Fatehi, Bingham, Dykeman, Stockley,
  and Twarock]{fatehi2021comparing}
F.~Fatehi, R.~J. Bingham, E.~C. Dykeman, P.~G. Stockley, and R.~Twarock.
\newblock Comparing antiviral strategies against {COVID}-19 via multiscale
  within-host modelling.
\newblock \emph{R. Soc. Open Sci.}, 8:\penalty0 210082, 2021{\natexlab{b}}.

\bibitem[Giachetti and Holland(1988)]{giachetti1988altered}
C.~Giachetti and J.~J. Holland.
\newblock Altered replicase specificity is responsible for resistance to
  defective interfering particle interference of an {S}di-mutant of vesicular
  stomatitis virus.
\newblock \emph{J. Virol.}, 62\penalty0 (10):\penalty0 3614--3621, 1988.

\bibitem[Gillespie(1977)]{gillespie1977exact}
D.~T. Gillespie.
\newblock Exact stochastic simulation of coupled chemical reactions.
\newblock \emph{J. Phys. Chem.}, 81\penalty0 (25):\penalty0 2340--2361, 1977.

\bibitem[Grow et~al.(2015)Grow, Flynn, Chavez, Bayless, Wossidlo, Wesche,
  Martin, Ware, Blish, Chang, et~al.]{grow2015intrinsic}
E.~J. Grow, R.~A. Flynn, S.~L. Chavez, N.~L. Bayless, M.~Wossidlo, D.~J.
  Wesche, L.~Martin, C.~B. Ware, C.~A. Blish, H.~Y. Chang, et~al.
\newblock Intrinsic retroviral reactivation in human preimplantation embryos
  and pluripotent cells.
\newblock \emph{Nature}, 522\penalty0 (7555):\penalty0 221, 2015.

\bibitem[Huang(1973)]{huang1973defective}
A.~S. Huang.
\newblock Defective interfering viruses.
\newblock \emph{Annu. Rev. Microbiol.}, 27\penalty0 (1):\penalty0 101--118,
  1973.

\bibitem[Huang and Baltimore(1970)]{huang1970defective}
A.~S. Huang and D.~Baltimore.
\newblock Defective viral particles and viral disease processes.
\newblock \emph{Nature}, 226\penalty0 (5243):\penalty0 325--327, 1970.

\bibitem[Kirkwood and Bangham(1994)]{kirkwood1994cycles}
T.~B.~L. Kirkwood and C.~R.~M. Bangham.
\newblock Cycles, chaos, and evolution in virus cultures: a model of defective
  interfering particles.
\newblock \emph{Proc. Natl. Acad. Sci.}, 91\penalty0 (18):\penalty0 8685--8689,
  1994.

\bibitem[Li and Aaskov(2014)]{li2014sub}
D.~Li and J.~Aaskov.
\newblock Sub-genomic {RNA} of defective interfering ({D.I.}) {Dengue} viral
  particles is replicated in the same manner as full length genomes.
\newblock \emph{Virology}, 468:\penalty0 248--255, 2014.

\bibitem[Li et~al.(2021)Li, Lin, Rawle, Jin, Wu, Wang, Lor, Hussain, Aaskov,
  and Harrich]{li2021dengue}
D.~Li, M.-H. Lin, D.~J. Rawle, H.~Jin, Z.~Wu, L.~Wang, M.~Lor, M.~Hussain,
  J.~Aaskov, and D.~Harrich.
\newblock Dengue virus-free defective interfering particles have potent and
  broad anti-dengue virus activity.
\newblock \emph{Commun. Biol.}, 4\penalty0 (1):\penalty0 557, 2021.

\bibitem[Lin and Lai(1993)]{lin1993deletion}
Y.~J. Lin and M.~M. Lai.
\newblock Deletion mapping of a mouse hepatitis virus defective interfering
  {RNA} reveals the requirement of an internal and discontiguous sequence for
  replication.
\newblock \emph{J. Virol.}, 67\penalty0 (10):\penalty0 6110--6118, 1993.

\bibitem[Mann et~al.(2006)Mann, Marriott, Balasingam, Lambkin, Oxford, and
  Dimmock]{mann2006interfering}
A.~Mann, A.~C. Marriott, S.~Balasingam, R.~Lambkin, J.~S. Oxford, and N.~J.
  Dimmock.
\newblock Interfering vaccine (defective interfering influenza {A} virus)
  protects ferrets from influenza, and allows them to develop solid immunity to
  reinfection.
\newblock \emph{Vaccine}, 24\penalty0 (20):\penalty0 4290--4296, 2006.

\bibitem[Manzoni and L{\'o}pez(2018)]{manzoni2018defective}
T.~B. Manzoni and C.~B. L{\'o}pez.
\newblock Defective (interfering) viral genomes re-explored: impact on
  antiviral immunity and virus persistence.
\newblock \emph{Future Virol.}, 13\penalty0 (7):\penalty0 493--503, 2018.

\bibitem[Marriott and Dimmock(2010)]{marriott2010defective}
A.~C. Marriott and N.~J. Dimmock.
\newblock Defective interfering viruses and their potential as antiviral
  agents.
\newblock \emph{Rev. Med. Virol.}, 20\penalty0 (1):\penalty0 51--62, 2010.

\bibitem[Metzger et~al.(2011)Metzger, Lloyd-Smith, and
  Weinberger]{metzger2011autonomous}
V.~T. Metzger, J.~O. Lloyd-Smith, and L.~S. Weinberger.
\newblock Autonomous targeting of infectious superspreaders using engineered
  transmissible therapies.
\newblock \emph{PLoS Comput. Biol.}, 7\penalty0 (3):\penalty0 e1002015, 2011.

\bibitem[Noble et~al.(2004)Noble, McLain, and Dimmock]{noble2004interfering}
S.~Noble, L.~McLain, and N.~J. Dimmock.
\newblock Interfering vaccine: a novel antiviral that converts a potentially
  virulent infection into one that is subclinical and immunizing.
\newblock \emph{Vaccine}, 22\penalty0 (23-24):\penalty0 3018--3025, 2004.

\bibitem[Notton et~al.(2014)Notton, Sardany{\'e}s, Weinberger, and
  Weinberger]{notton2014case}
T.~Notton, J.~Sardany{\'e}s, A.~D. Weinberger, and L.~S. Weinberger.
\newblock The case for transmissible antivirals to control population-wide
  infectious disease.
\newblock \emph{Trends Biotechnol.}, 32\penalty0 (8):\penalty0 400--405, 2014.

\bibitem[Pastuzyn et~al.(2018)Pastuzyn, Day, Kearns, Kyrke-Smith, Taibi,
  McCormick, Yoder, Belnap, Erlendsson, Morado, et~al.]{pastuzyn2018neuronal}
E.~D. Pastuzyn, C.~E. Day, R.~B. Kearns, M.~Kyrke-Smith, A.~V. Taibi,
  J.~McCormick, N.~Yoder, D.~M. Belnap, S.~Erlendsson, D.~R. Morado, et~al.
\newblock The neuronal gene {Arc} encodes a repurposed retrotransposon {Gag}
  protein that mediates intercellular {RNA} transfer.
\newblock \emph{Cell}, 172\penalty0 (1):\penalty0 275--288, 2018.

\bibitem[Patel et~al.(2017{\natexlab{a}})Patel, White, Thompson, Bingham,
  Wei{\ss}, Maskell, Zlotnick, Dykeman, Tuma, Twarock, et~al.]{patel2017hbv}
N.~Patel, S.~J. White, R.~F. Thompson, R.~J. Bingham, E.~U. Wei{\ss}, D.~P.
  Maskell, A.~Zlotnick, E.~C. Dykeman, R.~Tuma, R.~Twarock, et~al.
\newblock {HBV} {RNA} pre-genome encodes specific motifs that mediate
  interactions with the viral core protein that promote nucleocapsid assembly.
\newblock \emph{Nat. Microbiol.}, 2:\penalty0 17098, 2017{\natexlab{a}}.

\bibitem[Patel et~al.(2017{\natexlab{b}})Patel, Wroblewski, Leonov, Phillips,
  Tuma, Twarock, and Stockley]{patel2017rewriting}
N.~Patel, E.~Wroblewski, G.~Leonov, S.~E.~V. Phillips, R.~Tuma, R.~Twarock, and
  P.~G. Stockley.
\newblock Rewriting nature's assembly manual for a {ssRNA} virus.
\newblock \emph{Proc. Natl. Acad. Sci.}, 114\penalty0 (46):\penalty0
  12255--12260, 2017{\natexlab{b}}.

\bibitem[Poirier and Vignuzzi(2017)]{poirier2017virus}
E.~Z. Poirier and M.~Vignuzzi.
\newblock Virus population dynamics during infection.
\newblock \emph{Curr. Opin. Virol.}, 23:\penalty0 82--87, 2017.

\bibitem[Quintela et~al.(2018)Quintela, Conway, Hyman, Guedj, Dos~Santos,
  Lobosco, and Perelson]{quintela2018new}
B.~d.~M. Quintela, J.~M. Conway, J.~M. Hyman, J.~Guedj, R.~W. Dos~Santos,
  M.~Lobosco, and A.~S. Perelson.
\newblock A new age-structured multiscale model of the hepatitis {C} virus
  life-cycle during infection and therapy with direct-acting antiviral agents.
\newblock \emph{Front. Microbiol.}, 9:\penalty0 601, 2018.

\bibitem[Rand et~al.(2021)Rand, Kupke, Shkarlet, Hein, Hirsch,
  Marichal-Gallardo, Cicin-Sain, Reichl, and Bruder]{rand2021antiviral}
U.~Rand, S.~Y. Kupke, H.~Shkarlet, M.~D. Hein, T.~Hirsch, P.~Marichal-Gallardo,
  L.~Cicin-Sain, U.~Reichl, and D.~Bruder.
\newblock Antiviral activity of influenza {A} virus defective interfering
  particles against {SARS-CoV}-2 replication in vitro through stimulation of
  innate immunity.
\newblock \emph{Cells}, 10\penalty0 (7):\penalty0 1756, 2021.

\bibitem[Rast et~al.(2016)Rast, Rouzine, Rozhnova, Bishop, Weinberger, and
  Weinberger]{rast2016conflicting}
L.~I. Rast, I.~M. Rouzine, G.~Rozhnova, L.~Bishop, A.~D. Weinberger, and L.~S.
  Weinberger.
\newblock Conflicting selection pressures will constrain viral escape from
  interfering particles: Principles for designing resistance-proof antivirals.
\newblock \emph{PLoS Comput. Biol.}, 12\penalty0 (5):\penalty0 e1004799, 2016.

\bibitem[Rezelj et~al.(2018)Rezelj, Levi, and Vignuzzi]{rezelj2018defective}
V.~V. Rezelj, L.~I. Levi, and M.~Vignuzzi.
\newblock The defective component of viral populations.
\newblock \emph{Curr. Opin. Virol.}, 33:\penalty0 74--80, 2018.

\bibitem[Rezelj et~al.(2021)Rezelj, Carrau, Merwaiss, Levi, Erazo, Tran,
  Henrion-Lacritick, Gausson, Suzuki, Shengjuler, et~al.]{rezelj2021defective}
V.~V. Rezelj, L.~Carrau, F.~Merwaiss, L.~I. Levi, D.~Erazo, Q.~D. Tran,
  A.~Henrion-Lacritick, V.~Gausson, Y.~Suzuki, D.~Shengjuler, et~al.
\newblock Defective viral genomes as therapeutic interfering particles against
  flavivirus infection in mammalian and mosquito hosts.
\newblock \emph{Nat. Commun.}, 12\penalty0 (1):\penalty0 2290, 2021.

\bibitem[Rolfsson et~al.(2016)Rolfsson, Middleton, Manfield, White, Fan,
  Vaughan, Ranson, Dykeman, Twarock, Ford, et~al.]{rolfsson2016direct}
{\'O}.~Rolfsson, S.~Middleton, I.~W. Manfield, S.~J. White, B.~Fan, R.~Vaughan,
  N.~A. Ranson, E.~Dykeman, R.~Twarock, J.~Ford, et~al.
\newblock Direct evidence for packaging signal-mediated assembly of
  bacteriophage {MS2}.
\newblock \emph{J. Mol. Biol.}, 428\penalty0 (2):\penalty0 431--448, 2016.

\bibitem[Rouzine and Weinberger(2013)]{rouzine2013design}
I.~M. Rouzine and L.~S. Weinberger.
\newblock Design requirements for interfering particles to maintain coadaptive
  stability with {HIV}-1.
\newblock \emph{J. Virol.}, 87\penalty0 (4):\penalty0 2081--2093, 2013.

\bibitem[Shakeel et~al.(2017)Shakeel, Dykeman, White, Ora, Cockburn, Butcher,
  Stockley, and Twarock]{shakeel2017genomic}
S.~Shakeel, E.~C. Dykeman, S.~J. White, A.~Ora, J.~J.~B. Cockburn, S.~J.
  Butcher, P.~G. Stockley, and R.~Twarock.
\newblock Genomic {RNA} folding mediates assembly of human parechovirus.
\newblock \emph{Nat. Commun.}, 8\penalty0 (1):\penalty0 5, 2017.

\bibitem[Shirogane et~al.(2021)Shirogane, Rousseau, Voznica, Xiao, Su,
  Catching, Whitfiled, Rouzine, Bianco, and Andino]{Andino:2019}
Y.~Shirogane, E.~Rousseau, J.~Voznica, Y.~Xiao, W.~Su, A.~Catching, Z.~J.
  Whitfiled, I.~M. Rouzine, S.~Bianco, and R.~Andino.
\newblock Experimental and mathematical insights on the interactions between
  poliovirus and a defective interfering genome.
\newblock \emph{PLoS Pathog.}, 17\penalty0 (9):\penalty0 e1009277, 2021.

\bibitem[Stewart et~al.(2016)Stewart, Bingham, White, Dykeman, Zothner, Tuplin,
  Stockley, Twarock, and Harris]{stewart2016identification}
H.~Stewart, R.~J. Bingham, S.~J. White, E.~C. Dykeman, C.~Zothner, A.~K.
  Tuplin, P.~G. Stockley, R.~Twarock, and M.~Harris.
\newblock Identification of novel {RNA} secondary structures within the
  hepatitis {C} virus genome reveals a cooperative involvement in genome
  packaging.
\newblock \emph{Sci. Rep.}, 6:\penalty0 22952, 2016.

\bibitem[Stockley et~al.(2013{\natexlab{a}})Stockley, Ranson, and
  Twarock]{stockley2013new}
P.~G. Stockley, N.~A. Ranson, and R.~Twarock.
\newblock A new paradigm for the roles of the genome in {ssRNA} viruses.
\newblock \emph{Future Virol.}, 8\penalty0 (6):\penalty0 531--543,
  2013{\natexlab{a}}.

\bibitem[Stockley et~al.(2013{\natexlab{b}})Stockley, Twarock, Bakker, Barker,
  Borodavka, Dykeman, Ford, Pearson, Phillips, Ranson,
  et~al.]{stockley2013packaging}
P.~G. Stockley, R.~Twarock, S.~E. Bakker, A.~M. Barker, A.~Borodavka,
  E.~Dykeman, R.~J. Ford, A.~R. Pearson, S.~E.~V. Phillips, N.~A. Ranson,
  et~al.
\newblock Packaging signals in single-stranded {RNA} viruses: nature's
  alternative to a purely electrostatic assembly mechanism.
\newblock \emph{J. Biol. Phys.}, 39\penalty0 (2):\penalty0 277--287,
  2013{\natexlab{b}}.

\bibitem[Thompson and Yin(2010)]{thompson2010population}
K.~A.~S. Thompson and J.~Yin.
\newblock Population dynamics of an {RNA} virus and its defective interfering
  particles in passage cultures.
\newblock \emph{Virol. J.}, 7\penalty0 (1):\penalty0 257, 2010.

\bibitem[Thompson et~al.(2009)Thompson, Rempala, and Yin]{thompson2009multiple}
K.~A.~S. Thompson, G.~A. Rempala, and J.~Yin.
\newblock Multiple-hit inhibition of infection by defective interfering
  particles.
\newblock \emph{J. Gen. Virol.}, 90\penalty0 (4):\penalty0 888--899, 2009.

\bibitem[Thomson et~al.(1998)Thomson, White, and Dimmock]{thomson1998genomic}
M.~Thomson, C.~L. White, and N.~J. Dimmock.
\newblock {The genomic sequence of defective interfering Semliki Forest virus
  (SFV) determines its ability to be replicated in mouse brain and to protect
  against a lethal SFV infection in vivo}.
\newblock \emph{Virology}, 241\penalty0 (2):\penalty0 215--223, 1998.

\bibitem[Vignuzzi and L{\'o}pez(2019)]{vignuzzi2019defective}
M.~Vignuzzi and C.~B. L{\'o}pez.
\newblock Defective viral genomes are key drivers of the virus--host
  interaction.
\newblock \emph{Nat. Microbiol.}, 4\penalty0 (7):\penalty0 1075--1087, 2019.

\bibitem[von Magnus(1954)]{von1954incomplete}
P.~von Magnus.
\newblock Incomplete forms of influenza virus.
\newblock In \emph{Adv. Virus Res.}, volume~2, pages 59--79. Elsevier, 1954.

\bibitem[Yao et~al.(2021)Yao, Narayanan, Majowicz, Jose, and
  Archetti]{yao2021synthetic}
S.~Yao, A.~Narayanan, S.~A. Majowicz, J.~Jose, and M.~Archetti.
\newblock A synthetic defective interfering {SARS-CoV}-2.
\newblock \emph{PeerJ}, 9:\penalty0 e11686, 2021.

\bibitem[Yuan and Allen(2011)]{yuan2011stochastic}
Y.~Yuan and L.~J.~S. Allen.
\newblock Stochastic models for virus and immune system dynamics.
\newblock \emph{Math. Biosci.}, 234\penalty0 (2):\penalty0 84--94, 2011.

\bibitem[Zwart et~al.(2013)Zwart, Pijlman, Sardany{\'e}s, Duarte, Janu{\'a}rio,
  and Elena]{zwart2013complex}
M.~P. Zwart, G.~P. Pijlman, J.~Sardany{\'e}s, J.~Duarte, C.~Janu{\'a}rio, and
  S.~F. Elena.
\newblock Complex dynamics of defective interfering baculoviruses during serial
  passage in insect cells.
\newblock \emph{J. Biol. Phys.}, 39\penalty0 (2):\penalty0 327--342, 2013.

\end{thebibliography}

\newpage


{\centering
    \Huge\bf Supplementary material \\\vspace{2cm}
}

\renewcommand{\thefigure}{S\arabic{figure}}

\setcounter{figure}{0}

\begin{figure}[H]
	\centering
	\includegraphics[width=0.6\linewidth]{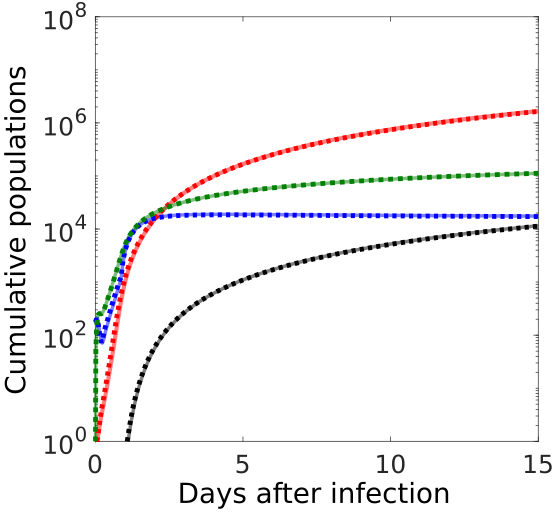}
	\caption{Comparing of the mean over 250 simulations of the Gillespie algorithm with the ODE system outcomes. Solid and dotted curves indicate the outcomes of the ODE system and Gillespie algorithm, respectively. The black, red, green and blue show the number of released virions, HCV RNAs, structural proteins and NS3 proteins. This figure shows that the mean over the Gillespie algorithm is same as the ODE model.}
	\label{FigS1}
\end{figure}


\end{document}